\documentclass[12pt]{article}

\usepackage{graphicx}

\title{Boosting QED and QCD  bound states in the path integral formalism
}
\author{  Yu.A.Simonov \\
State Research
Center\\Institute of Theoretical and Experimental Physics, \\
Moscow, 117218 Russia}
\newcommand{\beq}{\begin{eqnarray}}
 \newcommand{\eeq}{\end{eqnarray}}
\newcommand{\be}{\begin{equation}}
 \newcommand{\ee}{\end{equation}}

\def\fun#1#2{\lower3.6pt\vbox{\baselineskip0pt\lineskip.9pt
\ialign{$\mathsurround=0pt#1\hfil ##\hfil$\crcr#2\crcr\sim\crcr}}}
\newcommand{\veX}{\mbox{\boldmath${\rm X}$}}
\newcommand{{\SD}}{\rm SD}

\newcommand{{\Mc}}{\mathcal{M}}

\newcommand{\vex}{\mbox{\boldmath${\rm x}$}}
\newcommand{\vey}{\mbox{\boldmath${\rm y}$}}
\newcommand{\ver}{\mbox{\boldmath${\rm r}$}}

\newcommand{\veP}{\mbox{\boldmath${\rm P}$}}
\newcommand{\vep}{\mbox{\boldmath${\rm p}$}}
\newcommand{\veq}{\mbox{\boldmath${\rm q}$}}
\newcommand{\veQ}{\mbox{\boldmath${\rm Q}$}}
\newcommand{\vez}{\mbox{\boldmath${\rm z}$}}

\newcommand{\veR}{\mbox{\boldmath${\rm R}$}}

\newcommand{\vev}{\mbox{\boldmath${\rm v}$}}

\newcommand{\verho}{\mbox{\boldmath${\rm \rho}$}}

\newcommand{\vepi}{\mbox{\boldmath${\rm \pi}$}}

\newcommand{\llan}{\langle\langle}
\newcommand{\rran}{\rangle\rangle}
\newcommand{\lan}{\langle}
\newcommand{\ran}{\rangle}

\begin{document}
\maketitle
\begin{abstract}
Wave functions and energy eigenvalues of the path integral Hamiltonian are
studied in  Lorentz frame moving with velocity $v$. The instantaneous
interaction produced by the Wilson loop is shown to  be reduced by an overall
factor $\sqrt{1-(\frac{v}{c})^2}$. As a result one obtains the boosted energy
eigenvalues in the Lorentz covariant form $E= \sqrt{\veP^2+M^2_0}$, where $M_0$
is the c.m. energy, and this form is tested for two free particles and
  for the Coulomb and linear interaction.

 Using Lorentz contracted wave functions of the bound states one obtains the
 scaled parton wave functions  and valence quark distributions for large $P$. Matrix elements containing wave
 functions moving with different velocities strongly decrease with growing
 relative momentum, e.g. for the time-like formfactors one obtains $F_h
 (Q_0)\sim \left( \frac{M_h}{Q_0}\right)^{ 2 n_h}  $ with $n_h  = 1$ and 2 for
 mesons and baryons, as in the ``quark counting rule''.

\end{abstract}
\section{Introduction} Bound states in quantum field theories are usually considered in
the c.m. system, however many practical applications need the corresponding
wave functions in the arbitrarily moving system, e.g. in formfactors, reaction
etc.

On  the theoretical side the rules of the special relativity can predict the
frame dependence for the global characteristics,  such as the total energy.
However, for the internal characteristics  e.g.  wave functions, the general
theory is still unable to give answers, since the frame dependence of the full
dynamics includes the interaction term  \cite{1}.

A part of this difficulty is connected to the standard formalism of
relativistic bound state equations, based on the Bethe-Salpeter integral
equations, which is a complicated and not always  a rigorous instrument
(problems of the relative time, spurious solutions and difficulty to implement
nonperturbative effects).

The study of boost effects in this framework was done in \cite{2,3,4,5}, and
recently an analysis of boosted bound state equation for the $ep$ system was
presented in \cite{6}, while the $1+1 ~~e^+e^-$ case was considered in
\cite{7,8}.

Another approach (the $N$-quantum approach) was used in \cite{9}, while the WKB
method (expansion in powers of $\hbar$) in \cite{10}.

  A more convenient tool to study the effects of  boosting would be a
  Hamiltonian with e.g. instantaneous interaction, derived from the quantum
  field action. However this line of reasoning has  several complications.
  First of all, the  usual methods of introducing interaction via particle
  exchange - exclude the nonperturbative mechanisms, which produce confinement
  in QCD \cite{12}. Secondly, the Minkowskian particle  exchange automatically
  brings in the appearance  of  the  new Fock sectors  with  a new created
  particle in addition to the basic two-particle content. A way to handle this
  problem was suggested recently in \cite{7}. In a general case one obtains the
  matrix  Hamiltonian for the Fock column wavefunction, and one needs a
  systematic way to cut off the higher states to concentrate on the lowest
  ones.

  Moreover, one should have a simple way to disregard the spin degrees of
  freedom in the first approximation to make clear the dynamical mechanism of
  boosting.

  All these problems are economically treated in the path integral (world line)
  formalism \cite{13} based on the Fock-Feynman-Schwinger representation
  \cite{14}. Recently this formalism was presented in an especially useful form
  \cite{15}, where higher Fock state can be systematically eliminated, and the
  spin-dependent interaction treated separately \cite{16}. The accuracy of the
  first simple approximation in QCD and QED was checked in \cite{17}.

  As we show below, this formalism provides one with the Hamiltonian of an
  arbitrarily moving system with internal interaction, given by the Lorentz
  invariant construction -- the Wilson loop. The latter includes both
  nonperturbavive and perturbative  (particle exchange) interactions written
  first in the Euclidean space-time via field correlators and therefore does
  not produce higher Fock states, the latter appear due to perturbative quantum
  creation.

  Below we are using the path-integral formalism of \cite{15}, which provides
  the  Hamiltonian of a two-particle state in a moving coordinate system,
  where the total momentum $\veP$ of the state is  defined. Our formalism
  yields the Hamiltonian with  additional  parameters $\omega_1, \omega_2$,
  which play the role of  the  virtual particle energies, and the stationary
  point analysis of the state in terms of $\omega_1, \omega_2$,  exact  at
  large time interval of the system $T$, yields the energy $E$ of the moving
  state.

  It is  essential, that the interaction in the system is obtained in QCD and
  QED from  the Wilson loop, which is gauge and Lorentz invariant, and where
  the resulting instantaneous (or  light-cone) interaction is obtained via
  field correlators \cite{15}, see \cite{16*} for the corresponding light-cone
  work. In this way one can in principle define interaction in any moving
  system, however practically this task can be difficult. We are using general
  properties of the Wilson loop and make an ansatz for wave function. As  a
  result one can  find the approximate wave function and energy  of a given
  state in a moving coordinate system.

One should stress, that the study of the boosted dynamics by itself has not
much practical meaning, since all inertial systems are equivalent. However in
practice there are many processes, which include interactions of slow and  fast
particles, and, hence, possible overlap integrals of their wave  functions, and
here the knowledge of boosted dynamics is necessary, and as we show, it leads
to remarkable effects.

  In the present  paper we are addressing these problems for the QCD and QED
  bound states. We construct the Hamiltonian $H(\veP)$ of the two-body system
  and find its eigenvalues. We show that the eigenvalue in the moving system
  $E_0(\veP)$ can be found in the form
  \be E_0(\veP)= \sqrt{\veP^2 + \tilde M^2_0},\label{1m}\ee
  where $\tilde M_0$ is defined by the boosted interaction $V_{\veP}$ and
  should coincide with the c.m. energy $M_0$ provided $V_{\veP}$ is found
  correctly.

  Using the Wilson loop form, we find that the boost $L_{\veP}$ acts on the
  original c.m. interaction at least in two ways: reorganizing its magnitude
  $V\to CV$, and changing its coordinate dependence
  \be V_{\veP} = L_{\veP} V = CV (L \{\ver\})\label{2m}\ee

  From the general Lorentz invariance of the Wilson loop one finds that at the
  stationary points      \be C=C_0 = \sqrt{1-\left( \frac{v}{c}\right)^2 } =
  \frac{M_0}{E_0(\veP)}.\label{3m}\ee
  Note, that we shall put $c=1$ in what follows.
  Neglecting at first the second modification  $(V (\ver) \to V (L (\ver)))$,
  we check that the mass $\tilde M_0$ in (\ref{1m}) obtained in the moving
  system approximately coincides with the original mass eigenvalue calculated
  the c.m. system. We consider several examples of $V(\ver)$: 1) linear
  confinement, 2) gluon or photon exchange and in both cases we find that
  $\tilde M_0$ agrees with $M_0$ within 10\% or better.

  To check the full Lorentz covariance of our procedure with $\omega_1,
  \omega_2$ we consider the relativistic two-body system of noninteracting
  particles of different masses $m_1, m_2$ and apply our procedure.

  We find that indeed the eigenvalues $E_0(\vep)$ have the form (\ref{1m}), where $\tilde M_0$ is the c.m. of  two particles with the
  particle momenta $\vep_1=\vep, \vep_2=-\vep$ in the c.m. system, $\tilde M_0
  = M_0 (\vep, -\vep)$. Having checked this (global) invariance, one must
  consider the case  of the shape boosting  -- the so-called
  Lorentz--FitzGerald
  contraction \cite{17*}, which is the  result of the second    step in the
  boosting transformation $V(\ver) \to V (L (\ver)).$ This step cannot  be
  obtained directly and kinematically
  since the instantaneous interaction in one system is not Lorentz
  (kinematically) connected to the  instantaneous interaction in another
  moving system, and there occurs a serious  restructuring  of  the
  interaction.

  Therefore we impose the condition of the Lorentz contraction on the  wave function
  and interaction, and in this way we find the explicit form of $L(\ver)$ in
  (\ref{2m}). This condition, which plays the  role of an ansatz,  defines the
  wave function in any system, provided it is known in the c.m. frame. As a
  result one can connect the Lorentz  contracted wave function to the parton
  wave function at large $P$ and find,  that it has usual scaling properties,
  but interconnected transverse  and longitudinal dynamics. One also finds
  immediate consequences of Lorentz contraction for physical processes, e.g.
  for decay amplitudes, quark decay constants, form factors etc., part of which
  was discussed  before (see  e.g. \cite{2} and \cite{17**}), while other should be
  investigated in future.

  The paper is organized as follows. In the next section the general form of
  the path integral for two-body Green's function is written and the
  relativistic Hamiltonian is defined in terms of $\omega_1, \omega_2$ and
  Wilson loop interaction.

  In section 3 the case of the  two noninteracting particles is considered and
  the  general structure of Lorentz covariant eigenvalues $E_n=\sqrt{\veP^2 +
  M^2_n}$ is confirmed.

  In section 4 the global property of the boosted interaction $V\to CV, ~~ C=
  \frac{M_n}{E_n}$ is  derived from the Wilson loop  interaction and checked in
  the cases of Coulomb and confining interaction.

  The section 5 is devoted to the  Lorentz contraction ansatz and the ensuing
  properties of the wave function and interaction with possible consequences
  for physical effects. An appendix is devoted to the comparison of Lorentz
  contracted and scaled parton wave functions.

  In the last section the main results and approximations are reiterated and
  some prospectives are given.

\section{The QCD and QED path-integral Hamiltonian in  the moving  systems.}

The two-body $(q_1, \bar q_2)$ Green's function generated by the currents $j_i
= (\psi \Gamma_i \psi)$  in the relativistic path formalism \cite{15} can be
written in the Euclidean space-time \be G_{12} (x,y) = \frac{T}{2\pi}
\int^\infty_0 \frac{ d \omega_1}{\omega_1^{3/2}} \int^\infty_0
\frac{d\omega_2}{\omega_2^{3/2}} (D^3z^{(1)} D^3z^{(2)})_{\vex\vey}\lan \hat T
W_\sigma (A)\ran\label{2.1}\ee where $T$ is the Euclidean time interval, and
\be \lan \hat T W \ran = 4 tr Y \lan W \ran \exp (-K (\omega_1) - K(\omega_2)
), \label{2.2}\ee

\be Y= \frac14 \Gamma_1 (m_1 - i \hat p_1) \Gamma_2 (m_2 - i \hat
p_2)\label{m6}\ee and \be K(\omega) = \int^T_0 dt_E \left( \frac{\omega}{2} +
\frac{m^2}{2\omega} + \frac{\omega}{2} \left( \frac{ d\vez}{dt_E} \right)^2
\right), \label{2.3} \ee while $\lan W \ran $ is the Wilson loop, which can be
expressed via field correlators \be \lan W(C) \ran = \lan Tr \exp ig \int d
\pi_{\mu\nu} (z) F_{\mu\nu} (z) \ran= \exp \sum^\infty_{n=1} \frac{(ig)^n}{n!}
\int d\pi (1) ... d\pi(n) \llan F(1) ... F(n)\rran\label{2.4}\ee \be d\pi (i) =
d s_{\mu\nu} (z_{ i }) + i \sigma_{\mu\nu}
\frac{dt_i}{2\omega_i}.\label{2.5}\ee

As was shown in \cite{12} the leading quadratic in $F$ term in the cumulant
expansion (\ref{2.4}) in the c.m. frame yields  instantaneous confinement  and
gluon exchange interaction in the $3+1$ QCD case, and the Coulomb interaction
in $3+1$ QED

\be \lan W (C) \ran =\exp \left\{- \int V (r (t_E) ) dt_E+ {\rm spin~ dep.~
terms}\right\}\label{2.6}\ee \be V_{QCD} (r) = \sigma r -\frac43
\frac{\alpha_s}{r}, ~~ V_{QED} (r) =- \frac{\alpha}{r}.\label{2.7}\ee

It is clear that the vacuum averaging of $W(C)$, implied by the angular
brackets, yields higher order corrections to $\alpha_s,e^2$ and $ \sigma$.

We now turn to the case of the Hamiltonian in a moving Lorentz system. To this
end we define the $q_1\bar q_2$ Green's function and Hamiltonian in the system
with the total momentum $\veP$ \cite{15} $$G_P (x,y) = \frac{T}{2\pi}
\int^\infty_0 \frac{d\omega_1}{\omega_1^{3/2} }\int^\infty_0
\frac{d\omega_2}{\omega_2^{3/2}} Y d^3 (\vex-\vey)
e^{i\veP(\vex-\vey)}\times$$\be \times\lan \vex |e^{-H(\omega_1,\omega_2,
\vep_1, \vep_2)T}|\vey\ran\label{2.15}\ee with  \be H(\omega_1, \omega_2,
\vep_1,\vep_2) = \sum_i \frac{p^2_i + m^2_i + \omega^2_i}{2\omega_i} + \hat
V=\sum_{i=1,2} \frac{m^2_i + \omega^2_i}{2\omega_i} + \frac{\vepi^2}{2\tilde
\omega} + \frac{\veP^2}{2(\omega_1 + \omega_2)} + \hat V.\label{2.16}\ee Here
we have replaced the instantaneous c.m. coordinates $\vex_1$ and $\vex_2$, by
the relative and total c.m. coordinates \be \verho= \vex_1 -\vex_2, ~~\veR =
\frac{\omega_1 \vex_1 +\omega_2 \vex_2}{\omega_1+\omega_2},\label{2.17}\ee
$$\vepi = \frac{\partial}{i \partial \verho}, ~~\veP = \frac{\partial}{i\partial
\veR}.$$

Note, that $\vepi$ is not the c.m. momentum, the latter we denote for two
particles as $(\vep, -\vep)$.

 To have the lowest eigenvalue of $H(\omega_1,
\omega_2, \vep_1, \vep_2)$ one should calculate the minimal value of its
eigenvalue $E(\omega_1 , \omega_2, \veP)$ in the stationary value analysis with
respect to $\omega_1, \omega_2$ \be \left.\frac{\partial E (\omega_1, \omega_2
, \veP)}{\partial \omega_i}\right|_{\omega_i =\omega_i^{(0)}} = 0,~~ i=1,2; ~~
E_0 (\veP) \equiv E (\omega_1^{(0)} , \omega_2^{(0)}, \veP)\label{2.18}\ee

It is clear, that the instantaneous c.m. interaction $V_{cm}$ in (\ref{2.16})
will be changed by the Poincare boost  operator $\hat L(\veP)$ in the general
system, hence one can write for $\hat V$ in (\ref{2.16})\be\hat V = \hat L
(\veP) V_{cm}.\label{2.19}\ee

We expect that the eigenvalues of the operator (\ref{2.16}) after minimization
should have the form \be E_0 (\veP) = \sqrt{\veP^2 + M^2_0}\label{2.20}\ee
where $M_0$ is the c.m. energy of the $q_1\bar q_2$ system.

Note, that we have  neglected for simplicity spin-dependent forces originating
from $\sigma_{\mu\nu} \frac{dt}{2\omega}$ in (\ref{2.5}), see \cite{16} for
details and references. As a result our Hamiltonian is a $4\times 4$ unit
matrix in Dirac  indices. We shall check (\ref{2.20}) in the following chapters
in the examples of free particles and QCD and QED systems.

\section{The case of two noninteracting particles}

    In this case we define:
    \be \Omega\equiv \omega_1+\omega_2; ~~ \omega_1 =\Omega x, ~~\omega_2 =
    \Omega (1-x)\label{2.21}\ee
    \be H= \frac{\veP^2 +\Omega^2}{2\Omega}+ \frac{1}{2\Omega} M^2_0, ~~ M_0^2
    \equiv \frac{\vepi^2 + m^2_1}{x}+ \frac{\vepi^2+m^2_2}{1-x}.\label{2.22}\ee
    Minimizing $M_0^2$ in  the c.m. system, where $\vepi=\vep$ in the interval $x\in$[0,1] one obtains the stationary
    point $x_0 = \frac{\varepsilon_1}{\varepsilon_1+\varepsilon_2}$, where
    $\varepsilon_i = \sqrt{\vep^2+m^2_i}$, and  finally$M_0^2=
    (\varepsilon_1+\varepsilon_2)^2,$ so that minimizing finally in $\Omega$,
    one  has
    \be E_0 (\veP)=
    \sqrt{\veP^2+(\varepsilon_1+\varepsilon_2)^2}.\label{2.23}\ee

    Hence $\varepsilon_1 +\varepsilon_2$ is the c.m.energy of two free particles
    with masses $(m_1, m_2)$ and momenta $( \vep ,-\vep )$, and the Lorentz
    covariance of our expression is proved, if one assumes, that $M^2_0$ should  be  calculated in the c.m. system, where $\vepi =
    \vep$.Now one has to find the same  operator $M^2_0$ (\ref{2.22}) in the
    moving system.

We consider again  the Hamiltonian  for the free particle  motion, but now use
Lorentz transformations for momenta to  support the result  (\ref{2.23}) \be
H_0 = \frac{\veP^2}{2(\omega_1+\omega_2)} + \frac{\omega_1 + \omega_2}{2} +
\frac{\vepi^2}{2\tilde \omega} + \frac{m^2_1}{2\omega_1} +
\frac{m^2_2}{2\omega_2},\label{21m} \ee and all coordinates refer to the same
moment of time, $\tilde \omega =\frac{\omega_1\omega_2}{\omega_1+\omega_2}$.

In terms of individual momenta
 in the moving system  $\vep_i = \frac{1}{i} \frac{\partial}{\partial z_i}$,
 one can write
 \be \vep_1 = \frac{\omega_1}{\omega_1+\omega_2} \veP + \vepi, ~~ \vep_2 = \frac{\omega_2}{\omega_1+\omega_2} \veP -
 \vepi\label{22m}\ee
hence \be \veP = \vep_1 + \vep_2 , ~~ \vepi = \frac{\vep_1 \omega_2 - \vep_2
\omega_1}{\omega_1 + \omega_2}.\label{23m}\ee

Our next goal is to express $\vepi$ in terms of the relative c.m. momentum
$\vep  (c.m.) \equiv \vep$

Note, that the Hamiltonian (\ref{21m}) can be written as \be H_0 =
\frac{\vep^2_1+m^2_1 + \omega^2_1}{2\omega_1} + \frac{\vep^2_2 + m^2_2 +
\omega_2^2}{2\omega_2}\label{24v}\ee and the on-shell condition for $\omega_i$
is obtained from the stationary point analysis of (\ref{24v}) in terms of
$\omega_1$ and $\omega_2$. One has \be \omega_i^{(0)} = \sqrt{\vep^2_i +
m^2_i}, ~~ i=1,2.\label{25m}\ee

Hence $\omega_i^{(0)}$ plays the role of the particle energy and we shall below
always replace $\omega_i$ by  $\omega_i^{(0)}$, implying that both particles
are on the mass shell. We shall show that the eigenvalues of the   Hamiltonian
(\ref{21m}) satisfy Eq. (\ref{1m}) and $\tilde M_0$ is exactly equal to the
c.m. energy of two particles.

Introducing the velocity $v$ of the c.m. system, one can write \be \omega_1
=\frac{\bar \omega_1 + \vev\vep}{\sqrt{1-v^2}}, ~~ \omega_2 = \frac{\bar
\omega_2 - \vev\vep}{\sqrt{1-v^2}},\label{26m}\ee where  $\bar \omega_1$ are
particle energies in  the c.m. system. For parallel and perpendicular
components of $\vep_i$ w.r.t. $\vev$ in terms of the c.m. momenta $p_\|$ ,
$\vep_\bot$ one has \be p_1 (\|) = \frac{p_\|+v\bar\omega_1}{\sqrt{1-v^2}}, ~~
p_2 (\|) = \frac{-p_\|+v\bar\omega_2}{\sqrt{1-v^2}},\label{27m}\ee \be
p_1(\bot) = p_\bot, ~~ p_2 (\bot) = - p_\bot.\label{28m}\ee From (\ref{23m})
one obtains \be \vepi_\bot = \vep_\bot, ~~ \pi_\| = \frac{p_\|}{\sqrt{1-v^2}} +
\frac{v}{\sqrt{1-v^2}} \left( \frac{\omega_2 \bar \omega_1 - \omega_1 \bar
\omega_2}{\omega_1 + \omega_2}\right).\label{29m}\ee

Inserting (\ref{26m}) in  (\ref{29m}), one obtains \be \pi_\| = p_\|
\sqrt{1-v^2}\label{30m}\ee also $\omega_1  +\omega_2 = \frac{\bar \omega_1
+\bar \omega_2}{\sqrt{1-v^2}}$.

As a result $H_0$ assumes the form \be H_0 = \frac{\veP^2 \sqrt{1-v^2}}{2(\bar
\omega_1+ \bar \omega_2)} + \frac{\bar \omega_1 + \bar \omega_2}{2
\sqrt{1-v^2}} + \frac{\sqrt{1-v^2}}{2}\left \{ \frac{p^2_\| (1-v^2) + p^2_\bot
+ m^2_1}{\bar \omega_1 + \vev\vep} +\frac{p^2_\| (1-v^2) + p^2_\bot +
m^2_2}{\bar \omega_2 - \vev\vep} \right\}.\label{31m}\ee

Using $\bar \omega_i^2 = p^2_\| + p^2_\bot + m^2_i , ~~ i=1,2$ one finally
obtains \be H_0 = \frac{ \veP^2 \sqrt{1-v^2}}{2 (\bar \omega_1 + \bar
\omega_2)} + \frac{\bar \omega_1 + \bar \omega_2}{2 \sqrt{1- v^2}} + \frac{\bar
\omega_1 + \bar \omega_2}{2} \sqrt{1-v^2}.\label{32m}\ee One can express $v$ as
follows, $M_0 \equiv \bar \omega_1+\bar \omega_2$, \be v= \frac{P}{\sqrt{P^2
+M^2_0}}, ~~ \sqrt{1-v^2} = \frac{M_0}{\sqrt{P^2+M^2_0}}\label{33m}\ee and as a
result one has \be H_0 = \sqrt{\veP^2 + M^2_0} \label{33m*}\ee implying the
Lorentz covariance of our approach.

It is now easy to compare our results for the free relativistic particles with
that of the Newton-Wigner approach \cite{18}. there the c.m. relative momentum
$\vep$ and the total momentum $\veP$, defined as above in (\ref{22m}),
(\ref{23m}), are connected with the momentum $\vepi=\frac{1}{i}
\frac{\partial}{\partial\verho}$ as follows: \be \vep = \vepi + \frac{ (\vepi
\veP) \veP}{E_0 (\sqrt{\veP^2 + E^2_0 }+ E_0 )}.\label{34m}\ee

Here $\vep$ and $\veP$ are canonically conjugated to the relative and c.m.
coordinates  $\vex$ and $\veX$, defined as follows: \be \vex=\verho+ \frac{
(\verho \veP)\veP}{E_0 (\sqrt{\veP^2 + E^2_0 }+ E_0)} + \frac{ (\verho \veP)
\vep}{E_0 \bar \omega_1 \bar \omega_2} \left( \bar \omega_1 - \bar \omega_2 -
\frac{\vep \veP}{\sqrt{\veP^2 + E^2_0}}\right),\label{35m}\ee \be X_i = R_i -
\frac{S_{ik} P_k}{E_0 (\sqrt{\veP^2 + E^2_0 } + E_0)}\label{36m}\ee where \be
S_{ik} = \rho_i \pi_k - \rho_k \pi_i.\label{37}\ee One can use (\ref{33m}) to
reduce (\ref{34m}) to the simple result \be p_\| =\pi_\|
\frac{1}{\sqrt{1-v^2}}, ~~ p_\bot = \pi_\bot, \label{38m}\ee which coincides
with (\ref{29m}), (\ref{30m}). One can in principle express the total
Hamiltonian in terms of canonically conjugated coordinates $\vex, \veX$ and
momenta $\vep, \veP$ and proceed to solve with the interaction  taken into
account, which is a difficult road. Instead we shall investigate a simpler,
however approximate way,   where the boosted interaction is easier to
implement.

\section{Boosting QCD and QED interaction}

We return to the general form (\ref{2.16}) and discuss  the Lorentz boost of
the interaction. Note, that our result (\ref{2.16}) is obtained in the moving
Lorentz system. In particular, one can check, that $\int_C A_\mu dz_\mu$ as
well as $\int_S d s_{\mu\nu} F_{\mu\nu}$ is not only gauge, but also Lorentz
invariant structures.

Rewriting (\ref{2.6}) as \be \lan W\ran = \exp ( - i\bar V (\ver) T_M)
\label{39m}\ee one  can exploit the Einsteinian time  dilatation to predict,
that any averaged interaction obtained from the Wilson loop average, transforms
under Lorentz transformation  from system $K'$ to system $K {''}$, moving with
respect  to $K'$ with velocity $ v$ as follows \be\bar V' T' =  \bar V'' T'',
~~ T'' = \frac{T'}{\sqrt{1-v^2}}, ~~  V'' = \bar V'\sqrt{1-v^2}.\label{40m}\ee

This gives us the factor $C_0= \sqrt{1-v^2}$ announced in (\ref{3m}). However
in the off-shell form of the Hamiltonian (\ref{2.16}), before the stationary
analysis in $\Omega$ is done, we shall use the off shell form for  $C,
C=\frac{M_0}{\Omega}$, which reduces to $C_0$ at the stationary point
$\Omega=E_0$. Note, that  this is not the only effect of the system motion, and
we shall discuss the effect of the Lorentz contraction in section 5.

Let us now use the property (\ref{40m}) to calculate the mass $M^2_0$ for
different interactions. We shall write the Hamiltonian (\ref{2.16}) as \be H =
\frac{\veP^2}{2\Omega} + \frac{\Omega}{2} + \frac{\hat M^2_0}{2\Omega} , ~~
\frac{\hat M^2_0}{2\Omega} = \frac{\vepi^2}{2\tilde \omega} + C V (\vex) +
\frac{m_1}{2\omega_1}+ \frac{m^2_2}{2\omega_2}.\label{41m}\ee

We consider the cases of the Coulomb and linear interactions.

%%%%%%%%%%%%%%%%%%%%%%%%%%%%%%%%%%%%%%%%%
%%%%%%%%%%%%%%%%%%%%%%%%%%%%%%%%%%%%%%%%
%%%%%%%%%%%%%%%%%%%%%%%%%%%%%%%%%%

\begin{description}

    \item[1)]{\bf  The case of opposite charges in $\mathbf{3+1}$ QED}

    We now study eigenvalues of (\ref{2.16}), where $\hat V(\veP)$ is given by
    (\ref{2.19}) and we assume for simplicity, that $\hat L (\veP)$ may be
    represented by a simple factor, $\hat L(\veP)\to C (\veP)$.

    The eigenvalues of the equation
    \be\left(\frac{\vepi^2}{2\tilde \omega}- \frac{\alpha C}{r}\right)
    \varphi_n =\varepsilon_n\varphi_n\label{2.24}\ee are given by
    $\varepsilon_n =- \frac{\tilde \omega (\alpha C)^2}{2n^2}$ and the total
    energy is (notations are as in (\ref{2.21})),
    \be E(\omega_1, \omega_2, \veP) = \frac{\veP^2+\Omega^2}{2\Omega}+
    \sum_{i=1,2} \frac{m^2_i}{2\omega_i} - \frac{\tilde \omega
    (\alpha C)^2}{2n^2}.\label{2.25}\ee
    Writing $E(\omega_1,\omega_2,\veP)= E(\Omega, x,\veP)$, one obtains from
    the minimization in $x$,
    \be -\frac{m^2_1}{x^2} + \frac{m^2_2}{(1-x)^2} - \frac{(\Omega \alpha
    C)^2}{n^2} (1-2x) =0 .\label{2.26}\ee

For $m_1=m_2$ one obtains $x_0 =1/2, ~~ \omega_1^{(0)} = \omega_2^{(0)}\equiv
\omega_0$ \be E\left(\Omega, \frac12, \veP\right)=
\frac{\veP^2+\Omega^2}{2\Omega} + \frac{2m^2}{\Omega} - \frac{\Omega(\alpha
C)^2}{8n^2} \equiv \frac{\veP^2 +\Omega^2 +\mathcal{
M}^2_0}{2\Omega}.\label{2.26a}\ee Putting $C = \frac{\bar M}{\Omega}$, one
finds the minimum  of (\ref{2.26a}) varying in $\Omega$, which yields \be E_0
(\veP) = \sqrt{\veP^2+\mathcal{M}^2_0}, ~~ \Omega_0 =
\sqrt{P^2+\mathcal{M}_0^2},\label{2.27}\ee \be \mathcal{M}_0^2 =4m^2-
\frac{\alpha^2\bar M^2}{4n^2}.\label{2.28}\ee   Inserting $\bar
M=\mathcal{M}_0$ in (\ref{2.28}) one obtains \be \mathcal{M}_0 =
\frac{2m}{\sqrt{1+\frac{\alpha^2}{4n^2}}}.\label{2.29}\ee This is exactly the
spectrum found for the relativistic two-body Coulomb problem  neglecting spin
dependent and annihilation terms \cite{17}.

    \item[2)]{\bf The case of the linear confining potential in $\mathbf{3+1}$ QCD.}

    We are writing the Hamiltonian (\ref{2.16})  in the form

    \be H=\frac{\veP^2}{2\Omega} +\frac{\Omega}{2} + \frac{1}{2\Omega} \left(
    \frac{m_1^2}{x} + \frac{m^2_2}{1-x}+ \frac{\vepi^2}{x(1-x)} + 2
    \bar M  V(\ver)\right)\equiv \frac{\veP^2+\mathcal{M}^2_0}{2\Omega} +
    \frac{\Omega}{2}\label{2.30}\ee
    where we have denoted
    \be L(\veP)V_{cm} (\ver) =C V_{cm}
   (\ver),  ~~ C= \frac{\bar M}{\Omega},\label{2.31}\ee
and $\bar M$ can in principle depend on $\veP$ and $\vepi$, but we start with a
simple ansatz with $\bar M = const$.

For $m_1=m_2$ the minimization in $x$ yields $x_0 =\frac12 $ and \be
\mathcal{M}^2_0 = 4m^2 + \bar \varepsilon^2, ~~\varepsilon^2\varphi= (4\vepi^2
+ 2 \bar M  V(\ver))\varphi.\label{2.32}\ee For $V(\ver)=\sigma r$ the
eigenvalues of (\ref{2.32}) are known \cite{19}. \be \bar \varepsilon^2 =
(4\bar M\sigma)^{2/3} a_n, ~~ a_0 = 2.338, ~~ \mathcal{M}_0 = \sqrt{ 4m^2 + (4
\bar M \sigma)^{2/3} a_n}.\label{2.33}\ee Eqs. (\ref{2.32}), (\ref{2.33}) yield
solution for $\mathcal{M}_0 $ in the boosted system.

Let us now consider the massless case $m_1=m_2=0$ in (\ref{2.30}) and take
$\bar M  = \mathcal{M}_0 (c.m.)$, which we shall calculate below separately.
One has from (\ref{2.32}), (\ref{2.33}) \be \mathcal{M}_0^2 = (4\mathcal{M}_0
(c.m.) \sigma)^{2/3} a_n,\label{2.37}\ee while $ \mathcal{M}_0 (c.m.)$ is found
from (\ref{2.30})  with $\veP=0$ and $\bar M=\Omega$ \be H(c.m) =
\frac{\Omega}{2}+\frac{2\vepi^2}{\Omega}+\sigma r \to \mathcal{M}_0(c.m.) =
\frac{\Omega}{2} + \left( \frac{2}{\Omega}\right)^{1/3}\sigma^{2/3}
a_n.\label{2.38}\ee

Minimizing $\mathcal{M}_0 (c.m.)$ in $\Omega$ one obtains the standard
expression, commonly used in hadron spectra  \cite{19} \be \mathcal{M}_0 (c.m.)
= 4 \sqrt{\sigma} \left( \frac{a_n}{3}\right)^{3/4}.\label{2.39}\ee Insertion
of (\ref{2.39}) in (\ref{2.37}) immediately yields \be \mathcal{M}_0 = 3^{1/2}
(16)^{1/3} \sqrt{\sigma} \left( \frac{a_n}{3}\right)^{3/4} \approx 4.33
\sqrt{\sigma}  \left( \frac{a_n}{3}\right)^{3/4}.\label{2.40}\ee One can see,
that (\ref{2.39}) and (\ref{2.40}) coincide within the accuracy of 8\%, hence
one can use approximately the following boosting factor of the linear potential
\be L(\veP) V_{c.m.} (r) = \frac{\mathcal{M}_0 (c.m.)}{\Omega(\veP)} V_{c.m.}
(r).\label{2.41}\ee

\end{description}

\section{Lorentz contraction and the boosted shape of a  bound state in QCD and
QED}

The full (infinite) set of all paths for the two-body Green's function
(\ref{2.1}) can be studied by making intersection with the 3d hyperplane at
some fixed time $t_0$. In this way one obtains the density of the  state, given
by the wave function $\Psi_n$\be \rho_n (\vex, t) =\frac{1}{2 i} \left(\psi
\frac{\partial}{\partial t} \psi^+ - \psi^+\frac{\partial}{\partial t}
\psi\right) = E_n|\Psi_n (\vex, t)|^2.\label{57}\ee

For a boosted coordinate system  one can define the longitudinal and
perpendicular sizes of the system  $l_\|$ and $l_\bot$ \be l_\|= \int |x_\||
\rho_n (\vex, t) dV\label{58}\ee \be l_\bot= \int |\vex_\bot |\rho_n (\vex, t)
dV.\label{59}\ee

Since densities obey the invariance law \cite{17*}
 under Lorentz
transformations \be \rho (\vex, t ) d V = inv\label{60}\ee and $x_\|$ is
Lorentz contracted in a moving system, one expects the following  dependence of
$l_\|, l_\bot$ \be l_\| (v) = \sqrt{1-v^2} l_\| (0), ~~ l_\bot (v) = l_\bot
(0).\label{61}\ee

The  basic question is how   (\ref{61}) is maintained and supported by the
dynamics of the given system, i.e. what is the transformation law $L\{\ver\}$
in \be LV(\ver) = CV(L \{\ver\}).\label{62}\ee Instead of solving this point
for $V(\ver)$, we turn to the wave function  $\Psi_n (\vex, t)= e^{-i E_nt}
\varphi_n (\ver)$ and try to ensure the property (\ref{61}).

One  possible solution for $\varphi_n (\ver)$ to satisfy (\ref{58}), (\ref{59})
is \be L_P\varphi_n (r) =\varphi_n \left(x_\bot,
\frac{x_\|}{\sqrt{1-v^2}}\right)\label{63}\ee

On the other hand \be dV_v = (dx_\| d^2x_\bot )_v = \sqrt{1-v^2} (dx_\| d^2
x_\bot)_0\label{64}\ee so that Eq.(\ref{60}) is satisfied. Hence $$(l_\|)_v =
\int |x_\||E_n| \varphi_n \left( x_\bot, \frac{x_\|}{\sqrt{1-v^2}}\right) |^2
(dx_\| d^2 x_\bot)_v= $$\be= \sqrt{1-v^2} \int \tilde x_\|| M_0 |\varphi_n (
x_\bot, \tilde x_\|) |^2 d\tilde x_\| d^2 x_\bot = \sqrt{ 1- v^2}
(l_\|)_0,\label{65}\ee

where  we have used $\tilde x_\|= \frac{x_\|}{\sqrt{1-v^2}}=x_\|^{(0)}$, i.e.
the c.m. parallel coordinate.  The next question is what  kind of  dynamics can
ensure the transformation (\ref{63}).

We now turn to our Hamiltonian \be H=\frac{\veP^2}{2\Omega} +\frac{\Omega}{2} +
\frac{(h_0)_v}{2\Omega},\label{66}\ee where \be (h_0)_v = (4 \vepi^2_v+4 m^2 +
2 \bar M V_v), ~~V_v = LV_v, ~~\vepi^2_v = L \vepi^2\label{67}\ee

Note, that $\vepi$ is not canonical momentum and  in general the transformation
from the c.m. system with the instantaneous c.m. Hamiltonian $(h_0)_{cm}$ to
another instantaneous system with the total momentum $\veP$ is not provided by
Lorentz transformation. Therefore we start as before in section with the
assumption that $(h_0)_v$ should be first solved in the c.m. system, and this
result $(h_0)_{cm} \varphi_{cm} = M^2_0 \varphi_{cm}$ should be inserted as an
eigenvalue for $(h_0)_v$, i.e. $(h_0)_v \to (h_0)_{cm} = M^2_0$.

This assumption was proved to be true for the noninteracting pair in section 3
and approximately correct (within 10\%) for the typical interactions in section
4. We now generalize this assumption to specify  $\vepi^2_v$ and $V_v$, and to
this end we define \be (h_0)_v \varphi_v = M^2_0 \varphi_v , ~~ \varphi_v =
\varphi (x_\|^v, x_\bot^v)\label{68}\ee where \be x^v_\| =
\frac{x_\|}{\sqrt{1-v^2}}, ~~ x_\bot^v = x_\bot;~~ \pi^v_\|=\frac{1}{i}
\frac{\partial}{\partial x^v_\|},\label{69}\ee

\be \pi_\bot^2 = \frac{1}{i} \frac{\partial}{\partial x_\bot^v}, ~~ V_v =
V(x^v_\|, x^v_\bot).\label{70}\ee

As a result the equation $( h_0)_v \varphi_v = m_0^2\varphi_v$ reduces to the
c.m. equation $h_0 \varphi = M^2_0 \varphi$ by a simple change of variable
$x_\|$ with the same eigenvalue $M_0^2$. We now can write the boost
transformations for  $V$ and $\varphi_n$ as follows$$ V(r) =L V(X) \to CV
\left(x_\bot, \frac{x_\|}{\sqrt{1-v^2}}\right), ~~ C= \frac{M_0}{\Omega}
\to\sqrt{1-v^2},$$\be \varphi_n (x) \to \varphi_n^{(v)} (x) = \varphi_n \left(
x_\bot, \frac{x_\|}{\sqrt{1-v^2}}\right)\label{71}\ee with  the normalization
\be \int E_n |\varphi_n^{(v)} (x)|^2 d V_v =1= \int M_0^{(n)} | \varphi_n^{(0)}
(x) |^2 dV_0.\label{72}\ee

As a result the shape of the object will be transformed according to the
Lorentz contraction rule \be l_\| (v) = l_\bot (v) \sqrt{1-v^2}.\label{73}\ee

One of the immediate consequences of the form (\ref{72}) is the high momentum
asymptotics of the wave function \be \tilde \varphi_n^{(v)}(q) = \int
\varphi_n^{(v)} (\ver) e^{i\veq \ver} d^3 \ver = C_0 \tilde \varphi_0
(\veq_\bot, q_\| \sqrt{1-v^2}),\label{74}\ee where $\tilde
\varphi_0(q_\bot,q_\|)$ is the Fourier transformed c.m. wave function. One can
see in (\ref{74}), that in the case, when one of the constituents gets a large
momentum $\veQ(Q_\|,0,0)$, the boosted wave function has a  saturating limit up
to the factor $C_0$ namely
$$ \tilde \varphi_n^{(v)} (\veq + \veQ)= C_0 \tilde \varphi \left(\veq_\bot,
(q_\|+Q_\|) \sqrt{1-v^2_q}~\right)=$$\be
\begin{array}{c}{\longrightarrow}\\{{Q\to\infty}}\end{array} \frac{M_0}{Q} \tilde \varphi
\left(\veq_\bot, Q \frac{M_0}{\sqrt{Q^2+M^2_0}}\right) \to \frac{M_0}{Q} \tilde
\varphi (\veq_\bot, M_0)\label{75}\ee

As a second example we consider  the process of the strong  decay of a meson
$M_1$ in the state $n$ into a pair of hadrons $M_2, M_3$, which can be both
mesons or both baryons $B_2, B_3$. The corresponding amplitude can be written
for  the $B\bar B$ decay as \cite{20} $ ((\vep', \vep'')$ are two relative
momenta in the $3q$ system)\be J_{nB\bar B} (\vep) = \int y^{B\bar B}_{red}
\frac{d^3\vep'}{(2\pi)^3} \frac{d^3\vep''}{(2\pi)^3} \Psi^+_{n} (\vep-\vep'
-\vep'') \Psi_B (\vep', \vep'') \Psi_{\bar B} (-\vep', -\vep'')\label{76}\ee
and for the $M_2 M_3$ decay as \cite{21,22}.

\be J_{nn_2n_3} (\vep) = \int y^{M\bar M}_{red} \frac{d^3\veq}{(2\pi)^3}
\Psi^+_n (\vep + \veq) \Psi_{n_2} (\veq) \Psi_{n_3} ( \veq),\label{77}\ee where
$y_{red}$ contains the explicit decay parameters, which belong to the system at
rest of the initial meson, see \cite{20,21,22} for details.

It is important, that the boosted wave functions in both cases belong to the
decay products, moving with momenta $(\veP, - \veP)$, (where $M_1 = \sqrt{
M^2_2 + \veP^2} + \sqrt{M^2_3 + \veP^2},$) which grows as $M_1$, when $M_1$
lends to infinity.

As we have found in (\ref{74}), the momentum dependent wave function acquires
the factor $C_i= \sqrt{1-v^2_i}$ for each momentum, so that in (\ref{77})
$\psi_{n_i} (\veq) = C_i \varphi_{n_i} (\veq_\bot, q_\| \sqrt{1- v^2_i}),
i=2,3$ while in (\ref{76}) $$\Psi_B (\vep', \vep'') = (1- v^2_B) \varphi_B
(\vep_\bot', \vep_\bot'', p_\|'\sqrt{1-v^2_B}, p_\|''\sqrt{1-v^2_B})$$.

\begin{figure}[!htb]
\begin{center}
\includegraphics[angle=0,width=9 cm]{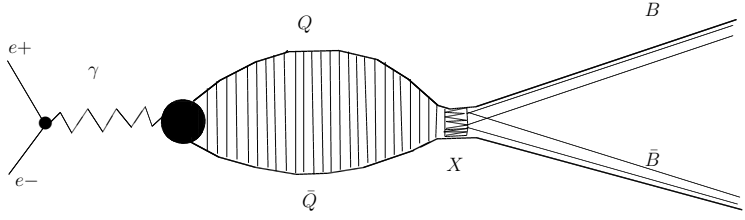}
\caption{The amplitude of the process $e^+e^- \to B\bar B$ with an intermediate
state $Q\bar Q(Q=u,d,s,...)$. The step $X$ in the process is a nonperturbative
double pair creation during a small time interval $\Delta t\sim \lambda\sim
0.1$ fm. }
\end{center}
\end{figure}

As a result $J_n$ behaves as \be J_{nB\bar B} \sim (1- v^2_B)^2\sim \left(
\frac{M^2_B}{M^2_B+\veP^2}\right)^2\label{78}\ee

 \be J_{nn_3n_3} \sim \sqrt{(1-v^2_2)
(1-v^2_3)} \sim \left(
\frac{M^2_2}{M^2_2+\vep^2}\frac{M^2_3}{M^2_3+\veP^2}\right)^{1/2}.\label{79}\ee

We can now turn to the processes $e^+e^-\to M_2 M_3$, $e^+e^- \to B\bar B$ and
the corresponding time-like formfactors, see Fig.1, and \cite{20} for details.

To this end we define the cross section \be \sigma (e^+e^- \to h \bar h) =
\frac{ 12 \pi \alpha^2 p}{E^3} \left| \sum_Q e_Q \sum_n \frac{\Psi_n (0)
\eta_{hQ} J_{nhh} (p)}{E_n - E + \frac{ i \Gamma_n}{2}} \right|^2,
\label{80}\ee where $h=M,B$ and $\eta_{hQ}$ is the spin-recoupling coefficient
and the time-like formfactor is proportional to (\ref{80}), $ F(Q_0) \sim
\frac{\sigma E^3}{p\alpha^24 \pi}$ where $Q_0 =E$

 Note, that the sum over $n$ with the energy
denominator has the same structure as in the total hadronic ratio $R$, and
brings in a constant up to possible logarithmic terms, see e.g.  \cite{24*} for
a discussion.

Therefore one obtains the following asymptotics of the formfactor, for mesons
and baryons  \be F_h (Q_0) \sim \left( \frac{M^2_h}{M^2_h + \vep^2}
\right)^{n_h}, ~~ h= M,B, n_M =1 , ~~ n_B=2.\label{81}\ee

Here $Q^2_0  = 4 (M^2_h + \vep^2)$. One can see, that the law (\ref{81})
coincides with the famous ``quark counting rule'' for  formfactors, found in
\cite{23}.

 Finally, one can connect our boosted wave function with the partonic function
 of the hadron, \cite{24}, which in the limit of high momentum $P$ can be
 written in the scaled form
 \be \Psi_N (\vep_1,...\vep_N)\to \psi (\{\vep_\bot^{(N)}\},
 x_1,...x_N)\label{79m}\ee
 where $x_i = \frac{p_\|^{(i)}}{P}$, and $\Psi$ is normalized in the Lorentz
 invariant way. We show below in the appendix, that for two  equal mass partons both
 functions are connected as (see Appendix for details)
 \be \psi_2 (\vep_\bot, x_1, x_2) = M_0 \sqrt{x_1x_2} (2\pi)^{-3/2} \tilde
 \varphi_0
 \left(p_\bot,  M_0\left( x_1 - \frac12\right)\right)\label{80m}\ee
 which implies, that $\tilde\varphi_0$ has a correct scaling limit and that transverse
 and longitudinal degrees of freedom are combined, e.g. for the $S$ wave one
 has $\tilde\varphi_0 = \varphi \left (\sqrt{ p^2_\bot + \left( x_1 - \frac12\right)^2 M^2_0}\right)$ and $M_0$ is the c.m.
 mass of the hadron.

 From(\ref{80m}) one can define the valence quark distribution function in the
 hadron  $D^q_h (x, k_\bot),$ e.g. for a meson
 \be D^q_h (x, k_\bot) = \frac{ E^2_h}{\varepsilon_1 \varepsilon_2} | \psi_2 (
 p_\bot, x) |^2 = \frac{M_0^2}{(2\pi)^3} \left|\tilde \varphi_0  \left(p_\bot,
 M_0\left(x-\frac12 \right)\right)\right|^2.\label{83}\ee

 Note, that (\ref{83}) does not show any singularities, as $\tilde\varphi_0$ is a
 monotonic Fourier transform of the static hadron wave function, and the
 corresponding quark distribution function $D^q_h (x) = \int d^2 k_\bot D^q_h
 (x, k_\bot)$ is decreasing for large $\left| x_1 - \frac12\right|$ and has a maximum at $x=\frac12$, and no peak around $x=0$.
  At this point one should remember,
 that we are treating the {\bf isolated} valence Fock component  without the multigluon Regge
 ladder Fock components, which bring in the expected singularity
  of $D^q_h(x)$ at $x$ small. Nevertheless $\tilde \varphi_0 (k_\bot, k_\|)$ can be a good
 starting approximation for the perturbative process evolution.

 One should stress again, that we have imposed the Lorentz contraction mechanism
 on the valence component only, while higher components ( wee partons) may
 create a more complicated picture.

 Therefore our valence component could be called the ``pure valence''
 component, which is different from the standard valence component , e.g. $u^v(x) =
 u(x) -\bar u(x)$, containing Regge ladder contribution,  and it has no singularities at small $x$.

 \section{Conclusion and prospectives}

 We have used path integral technic to calculate Hamiltonian and wave
 functions of the boosted system of two particles interacting via Wilson loop average. In  doing so we have neglected  all Fock
 components of the wave function except for the chosen $(q\bar q)$ or $(e_1
 e_2)$.

 This may be  called a quenched approximation, but inclusion of higher Fock
 components is technically possible, see  e.g. \cite{25}, \cite{26}. We have proceeded in
 in  the  following steps:

 1) We have used the Wilson loop interaction to prove, that  it  acquires due
 to the boost an overall reduction factor $C= \frac{M_0}{\Omega} \to \sqrt{1-v^2}$.

 2) We have  considered a free system of two relativistic particles with
 different masses $m_1, m_2$ and proved, that our Hamiltonian correctly
 describes  the spectrum  of the  boosted systems, when the relative motion
 part $(h_0)$ is computed in the c.m. system.

 3) Combining 1) and 2) we have  checked   that our boosted  Hamiltonian $H=
 \frac{\veP^2 + \Omega^2 + h_0}{2\Omega}, ~~ h_0 = h^{rel}_{kin} + 2 M_0 V$
 approximately (within 10\%) reproduces the  boosted spectrum
 $$ E(P) = \sqrt{\veP^2+ M^2_0}$$
 after the  minimization in $\Omega$ is imposed on $H$. This check was done for
 the Coulomb and confining interaction.

 4) We have used the Lorentz contraction phenomenon to formulate  the ``Lorentz
 contraction rule'', by which the two-body wave function in its
 dependence on the relative instantaneous coordinate $r$ in the moving system,
 behaves as $\varphi (\ver_\bot, \frac{r_\|}{\sqrt{1-v^2}})$, where $v$ is the
 velocity of the system.

 Correspondingly the Fourier transform of the wave function \\ $C\tilde \varphi
 (p_\bot, \sqrt{1-v^2} (p_\| +Q))$ appears to be sensitive to large momentum
 transfer $Q$,  only through the  factor $C_0=\sqrt{1-v^2}.$

 As a result we have shown, that the amplitude of the two-meson decay process
 with large decay energy is proportional to the product $[(1-v^2_1)
 (1-v^2_2)]^{1/2}$. Moreover, as a consequence, the time-like form factors of
 mesons and baryons behave as $\frac{1}{Q^2}$ and $\frac{1}{Q^4}$ respectively,
 i.e. the same way, as in the ``quark counting rule''.

 5) We have connected our Lorentz contracted wave function with the scaled
 partonic function and found a smooth limit between them. In this way one can
 insist, that our nonperturbative wave function can be used in hard processes,
 which opens a wide field  of applications. Note at this point, that the effect
 of Lorentz contracted wave functions was discussed before in the literature in
 different forms, see e.g. the first reference in \cite{2} for a discussion
 and additional references, and application to the reaction $\vep\bar{\vep} \to \pi\pi$
 in \cite{17**}. Within the framework of the ``Lorenz contraction rule'' the
 initial and  final wave functions entering in the hard   processes can be
 directly taken from the c.m. bound state wave function $\psi_{cm}$.  We also
 proved that the properly defined through $\psi_{cm}$   parton wave functions  have correct
 scaling limits not depending on the boost momentum $\veP$.

 The approach presented above,   calls for further study and applications in numerous
 processes, where boosted wave functions appear in initial and final states,
 e.g. in form factors, decay processes, etc.

 The author is grateful to B.L.Ioffe and O.V.Kancheli for useful discussions.
 This research was supported by the RFBR grant 1402-00395.

%\newpage
\vspace{2cm}
 \setcounter{equation}{0}
\renewcommand{\theequation}{A.\arabic{equation}}

\hfill {\it  Appendix  1}

\centerline{\it\large  Normalization of boosted many particle states}

 \vspace{1cm}

\setcounter{equation}{0} \def\theequation{A. \arabic{equation}}

We start with the normalization of the multiparton wave function \cite{24} \be
E(P) \int \prod^N_{i=1} \frac{d^3p_i}{\varepsilon_i} \delta \left(
P-\sum^n_{k=1} p_k\right)| \psi(p_1,... p_N)|^2 =1 , \varepsilon_i =
\sqrt{\vep^2_i + m^2_i}\label{A.1}\ee which in the limit of large $P$ for $x_k
= \frac{p_\|^{(k)}}{P}$ can be written as (fast moving partons) \be \int \prod
d^2 p_\bot^{(i)} \frac{dx_i}{x_i} \delta^{(2)} (\sum^N_{i=1} \vep_\bot^i)
\delta (1- \sum^N_{i=1} x_i) |\psi( p_\bot^{(i)} , x_i)|^2=1 .\label{A.2}\ee

For two partons our wave function (\ref{72}) is normalized as \be M_0 \int
|\tilde\varphi_0 (p_\bot, p_\|) | \frac{d^3p}{(2\pi)^3} =  E(P) ({1-v^2}) \int
|\tilde \varphi_0 (k_\bot, k_\| \sqrt{1-v^2}|^2 \frac{d^3k}{(2\pi)^3}
=1,\label{A.3}\ee and the parton wave function $\psi (p_1, p_2)$ in terms of
our $\varphi(k)$ is

\be \psi (p_1, p_2) = \sqrt{\frac{\varepsilon_1 \varepsilon_2
M_0^2}{E(P)(2\pi)^{3 }}  }  \tilde \varphi_0,  ~~ \sqrt{ 1-v^2} =
\frac{M_0}{\sqrt{P^2+M^2_0}}.\label{A.4}\ee

To express $k_\bot, k_\|$ via parton momenta $p_i$ one can use (\ref{27m});
hence \be p_\|^{(1)} = k_\|+ \frac{P}{M_0} \bar\omega_1= P x_1\label{a.5}\ee
where $M_0=\bar \omega_1 + \bar \omega_2$, (which is approximately true also
for interacting partons),  and for equal mass partons one has \be x_1 =
\frac{k_\|}{P} + \frac{\bar \omega_1}{M_0} \to \frac{k_\| \sqrt{1-v^2}}{M_0} +
\frac12.\label{a.6}\ee

 At large $P\gg M_0, $
  one has the scaling limit \be \psi(p_\bot^{(i)} , x_i) = \frac{M_0}{(2\pi)^{3/2}} \sqrt{x_1 x_2}
 \tilde \varphi_0 \left(p_\bot, x_1-\frac12\right).\label{A.5}\ee
In a similar way the multiparton wave function is connected to our wave
function $\varphi(\vepi_1,... \vepi_{N-1})$ as \be \psi_N (p_1, ... p_N) =
\frac{1}{ (2\pi)^{\frac32 (N-1)}} \prod^N_{i=1} \sqrt{M_0 x_i} \varphi_N
(p^{(i)}_\bot, M_0 (x_i- {\nu_i} )).\label{A.6}\ee One should  stress again,
that $\tilde \varphi_0$ has no on-shell singularities and there is no
factorization in $p_\bot^{(i)}, x_i,$ since at lowest angular momentum\\
$\tilde \varphi_0  \sim \tilde \varphi \left(  \sqrt{(p_\bot)^2 + M^2_0\left( x
- \frac12\right)^2}\right)$.


\begin{thebibliography}{99}
%

\bibitem{1}

P.A.M. Dirac, Rev. Mod. Phys. {\bf 21}, 392 (1949)

\bibitem{2}
Y.S.Kim and R.Zaou, Phys. Rev. {\bf D4}, 1764 (1971)\\
S.R.Coleman, Annals Phys. (NY), 101, 239 (1976).

\bibitem{3}
D.Alba, H.W.Crater and L.Lusanna, J.Phys. {\bf A40}, 9585 (2007).

\bibitem{4}
M.G.Rocha, F.J.Llanes-Estrada, arXiv:0910.1448; D.Schuette, S.Villaba-Chavez,
Eur. J. Phys. {\bf A44} (2010); W.N.Polyzou, arXiv:0908.1441.



\bibitem{5}
D.G.Currie, T.F.Jordan and E.C.G.Sudarshan, Rev.Mod. Phys. {\bf 35}, 350
(1963).

\bibitem{6} M.J\"{a}rvinen, Phys. Rev. {\bf D71}, 085006 (2005).



\bibitem{7} D.Dietrich, P.Hoyer and M.J\"arvinen, Phys. Rev. {\bf D85}, 1405016
(2012).




\bibitem{8} M.J\"{a}rvinen, Phys. Rev. {\bf D70}, 065014 (2004).

\bibitem{9} S.Cowen, arXiv:1107.0958.




\bibitem{10} S.J.Brodsky and P.Hoyer, Phys.Rev. {\bf D83}, 045026 (2011);
arXiv: 0909.3045

\bibitem{12} H.G.Dosch, Phys. Lett. {\bf B190}, 177 (1987);  H.G.Dosch and Yu.A.Simonov, Phys. Lett. {\bf B205}, 339
(1988);

Yu.A.Simonov, Nucl. Phys. {\bf B307}, 512 (1988); Phys. Usp.,{\bf 39}, 313
(1996).






\bibitem{13} R.P.Feynman, Rev. Mod. Phys. {\bf 20}, 367 (1948);\\ R.P.Feynman
and A.R.Hibbs, Quartum Mechanics and Path Integrals, (Mc Graw-Hill, New York,
1965);\\ Yu.A.Simonov, Nucl. Phys. {\bf B307}, 512 (1988).



\bibitem{14}  Yu.~A.~Simonov and J.A.Tjon, Ann. Phys. (N.Y) {\bf  228}, 1 (1993); ibid {\bf 300}, 54 (2002).


\bibitem{15}
 Yu.~A.~Simonov,
Phys. Rev.  {\bf  D88}, 025028 (2013); arXiv:1303.4952.




\bibitem{16}  Yu.~A.~Simonov, Phys.Rev.   {\bf D 88}, 053004 (2013);
arXiv:1304.0365.


\bibitem{17}  Yu.~A.~Simonov, Phys.Rev.   {\bf D 90}, 013013 (2014), arXiv: 1402.2162.


\bibitem{16*}
A.~Yu.~Dubin, A.~B.~Kaidalov, and Yu.~A.~Simonov, Phys.Lett. {\bf B343}, 310
(1995);
 V.L.Morgunov, V.I.Shevchenko and Yu.A.Simonov, Phys. Lett. {\bf B 416}, 433
 (1998).


\bibitem{17*}H.A.Lorentz, Encyklop\"{a}die der Mathematischen Wissenschaften, Band V, art. 13,14, Leipzig,
1904;\\ W.Pauli,  Theory of Relativity, Pergamon Press, N.Y., 1958.


\bibitem{17**}

B.El-Bennich and W.M.Kloet, Phys. Rev. {\bf C68}, 014003 (2003).


\bibitem{18} T.D.Newton and E.P.Winger, Rev.Mod. Phys. {\bf 21}, 400 (1949);\\
 Yu.~S.~Kalashnikova and  A.~V.~Nefediev,  Phys. At. Nucl.  {\bf  61}, 785
 (1998); ibid. {\bf 60}, 1389 (1997).


\bibitem{19} E.Eichten, K.Gottfried, T.Kinoshita, K.D.Lane and T.M.Yan, Phys. Rev. {\bf D 17}, 3090 (1978); ibid {\bf D21}, 203 (1980.
Yu.A.Simonov in: QCD: Perturbative or Nonperturbative?, L.Ferreira, P.Nogueira
and J.Silva-Marcos eds., Interscience, Singapore, 2000; hep-ph/9911237.


\bibitem{20} Yu.~A.~Simonov, Phys. Rev.  D {\bf 85}, 105025 (2012),  ibid. 125025 (2012), arXiv:1109.5545.


\bibitem{21} V.D.Orlovsky and Yu.A.Simonov, Phys. Rev. {\bf D 84}, 065013
(2011).


\bibitem{22} A.~M.~Badalian V.D.Orlovsky and Yu.A.Simonov, Phys. At. Nucl. {\bf
76}, 525 (2013).

\bibitem{24*} Yu.A.Simonov, Phys. At. Nucl. {\bf 65}, 135 (2002);
hep-ph/0109081; Phys. At. Nucl. {\bf 58}, 107 (1993), hep-ph/9311247.



\bibitem{23} V.A.Matveev, R.M.Muradyan and A.N.Tavkhelidze, Lett. Nuovo Cim. {\bf 7}, 719
(1973);\\
S.J.Brodsky and G.R.Farrar, Phys. Rev. Lett. {\bf 31}, 1153 (1973).

\bibitem{24} R.P.Feynman, Photon-Hadron Interactions, W.A.Benjamin Inc. New
York, 1972;\\
B.L.Ioffe, V.A.Khose, and L.N.Lipatov, Deep Inelastic Processes, North-Holland,
1984;\\ F.J.Yndurain, The Theory of Quark and Gluon Interactions, Springer,
2006.

\bibitem{25} A.B.Kaidalov, At the Frontier of Particle Physics, hep-ph/0103011;
M.Shifman ed. vol.1,603;\\ F.Gross, I.V.Musatov, Yu.A.Simonov, Phys. At. Nucl.
{\bf 69}, 699 (2006).
\bibitem{26}Yu.A.Simonov, Phys. At. Nucl. {\bf 67}, 553 (2004);
hep-ph/0306310;\\
Yu.A.Simonov, in `` I.Ya.Pomeranchuck and physics at the turn of the Century'',
A.Berkov, N.Narozhny and L.Okun eds. World Scientific, Singapore, 2003;
hep-ph/0310031.




\end{thebibliography}
\end{document}